\documentclass{aa}

\begin{document}

\thesaurus{06            % A&A Section 6 : Form. struct. & evol. of stars
	       (08.14.2;          % novae, cataclysmic variables
 		13.25.5;          % X-rays: stars
		08.13.1;          % Stars: magnetic fields
		02.01.2;          % Accretion, accretion disks
		08.09.02 YY Dra;  % Stars: individual: YY Dra
		08.09.02 V709 Cas % Stars: individual: V709 Cas
		)}

\title{YY~Draconis and V709~Cassiopeiae: two intermediate polars 
       with weak magnetic fields}

\titlerunning{YY~Dra and V709~Cas}

\author{A.J. Norton\inst{1} \and A.P. Beardmore\inst{2} \and 
     Alasdair Allan\inst{2} \and Coel Hellier\inst2}

\institute{Department of Physics, The Open University, Walton Hall, 
        Milton Keynes MK7 6AA \and 
	Department of Physics, Keele University, Keele,
	Staffordshire ST5 5BG}

\authorrunning{Norton et al.}

\date{Received ?? ??? 1998 /
      Accepted ?? ??? ????}

\maketitle

\begin{abstract}

We present data from long {\em ROSAT} HRI observations of the intermediate
polars \object{YY Dra} and \object{V709 Cas} which show that V709~Cas, 
like YY~Dra, exhibits a double-peaked X-ray pulse profile. Neither system 
shows evidence for X-ray beat period or orbital modulation, so both must be 
disc-fed accretors seen at low inclination angles. We argue that 
the short spin periods of the white dwarfs in these objects indicate that 
they have weak magnetic fields, so the radius at which material is captured 
by the field lines is relatively small. Consequently the footprints of the 
disc-fed accretion curtains on the white dwarf surface are large. 
The optical depths to X-ray emission within the accretion curtains are 
therefore lowest in the direction along the magnetic field lines, and highest 
in the direction parallel to the white dwarf surface, such that the emission 
from the two poles conspires to produce double-peaked X-ray pulse profiles. 
We emphasise that such a pulse profile is {\em not} a unique indicator of 
two-pole accretion however. Indeed, two-pole accretion onto smaller regions 
of the white dwarf surface may be considered the `normal' mode of behaviour 
in a disc-fed intermediate polar with a longer white dwarf spin period (and 
therefore a higher field strength), resulting in a single-peaked pulse 
profile.

Collating data on other intermediate polars, we may classify them into 
two subsets. Fast rotators, with relatively weak fields, show double-peaked 
pulse profiles (\object{AE Aqr}, \object{DQ Her}, \object{XY Ari}, 
\object{GK Per}, V709~Cas, YY~Dra, \object{V405 Aur}), whilst slower 
rotators, with larger fields and therefore larger magnetospheres, have been 
seen to exhibit an X-ray beat period modulation at some time (\object{FO Aqr},
\object{TX Col}, \object{BG CMi}, \object{AO Psc}, \object{V1223 Sgr}, 
\object{RX J1712.6--2414}).


\end{abstract}

\section{Introduction}

Intermediate polars are semi-detached interacting binaries in which a
magnetic white dwarf accretes material from a Roche-lobe filling, usually
late-type, main sequence companion star. The accretion flow from the
secondary proceeds towards the white dwarf either through an accretion
disc, an accretion stream, or some combination of both (known as disc 
overflow accretion), 
until it reaches the magnetospheric radius. Here the material attaches to the
magnetic field lines and follows them towards the magnetic poles of the 
white dwarf. The infalling material that originates from an accretion disc 
takes the form of arc-shaped accretion curtains, standing above the white 
dwarf surface. At some distance from this surface, the accretion flow 
undergoes a strong shock, below which material settles onto the white dwarf, 
releasing X-rays as it cools by thermal bremsstrahlung processes. Since the 
magnetic axis is offset from the spin axis of the white dwarf, this gives 
rise to the defining characteristic of the class, namely X-ray emission pulsed 
at the white dwarf spin period. If any of the material accretes directly from 
an accretion stream, the proportion falling onto each pole of the white 
dwarf will vary according to the rotation phase of the white dwarf in the 
reference frame of the binary. Consequently, stream-fed (or disc overflow) 
accretion will give rise to X-ray emission that varies with the beat 
period, where $1/P_{\rm beat} = 1/P_{\rm spin} - 1/P_{\rm orbit}$. About 
twenty confirmed intermediate polars are now recognized with a similar 
number of candidate systems having been proposed. Comprehensive reviews of 
various aspects of their behaviour are given by Patterson (\cite{Patt}), 
Warner (\cite{War95}), Hellier (\cite{Hell95}; \cite{Hell96}) and Norton 
(\cite{Nor2}).

\section{Observational histories}

\subsection{YY~Draconis}

The 16th magnitude star YY~Dra was discovered in 1934 and originally 
mis-classified as an Algol-like system. Detected as an X-ray source by 
{\em Ariel V} (3A~1148+719) and {\em Einstein} (2E~1140.7+7158), it was 
subsequently reclassified as a cataclysmic variable (see the discussions in 
Patterson et al. \cite{P92} and Patterson \& Szkody \cite{PS} for the history 
of this source). Optical radial velocity observations by Friend et al. 
(\cite{Friend}) revealed its orbital period as 3.97~hr, whilst time resolved 
infrared spectroscopy by Mateo, Szkody \& Garnavich (\cite{Mateo}) yielded 
estimates of other system parameters. The intermediate polar nature of YY~Dra 
was eventually recognised by Patterson et al. (\cite{P92}), who discovered 
optical photometric modulation at periods of about 265~s and 275~s. 
Indications of sub-harmonics 
at periods of 529~s and 550~s suggested that the spin period of the white 
dwarf might actually be 529~s, and that 550~s was the beat period between the 
orbital and spin periods (ie. the spin period of the white dwarf in the 
binary reference frame). The dominant signals at the shorter periods 
indicated that the pulse profile is `double-peaked'.

A series of short X-ray observations of YY~Dra with the {\em ROSAT} PSPC and 
HRI (Patterson \& Szkody {\cite{PS}; Beuermann \& Thomas \cite{beuer}; 
Reinsch et al. \cite{Reinsch95}) confirmed the earlier findings and revealed 
the following periods: 529.2~s (1990, PSPC all sky survey); 265~s (1991, 
PSPC); 264~s (1992, HRI); 264.6~s and 530.9~s (1993, PSPC). These authors
concluded that the true spin period of the white dwarf was around 530~s, 
that the presence of a dominant signal at half this period indicates 
that the system usually accretes onto both poles of the white dwarf, and 
that the emission sites are nearly identical.

Subsequent Hubble Space Telescope spectroscopy of YY~Dra (Haswell
et al. \cite{Has97}) established a more accurate white dwarf spin period of 
529.31~s, by combining their detection of UV pulsations at a period 
of 264.71~s with the previous detections referred to above.
A beat period of 273~s was also detected and attributed to reprocessing 
of the X-ray pulse in a structure fixed in the orbital frame. Haswell 
et al. determined a precise orbital period of 3.968976~hr and calculated 
the system parameters, including an inclination angle of 45$^{\circ}$. 

\subsection{V709~Cassiopeiae}

This source was recognised as an intermediate polar by Haberl \& Motch 
(\cite{HM95}), following its detection in the {\em ROSAT} \ All Sky Survey 
as RX~J0028.8+5917. A follow up 18~ksec pointed observation with the 
{\em ROSAT} \ PSPC revealed a pulse period of 312.8~s and a conventional 
`hard' intermediate polar X-ray spectrum. Motch et al. (\cite{Motch}) 
subsequently noted that RX~J0028.8+5917 was probably coincident with 
previously catalogued sources detected by {\em HEAO-1} \ (1H~0025+588), 
{\em Uhuru} \ (4U~0027+59) and {\em Ariel~V} \ (3A~0026+593), and identified 
the X-ray source with a 14th magnitude blue star, V709~Cas. The optical 
spectra of this star show radial velocity variations with periods of either
 5.4~h or 4.45~h, the two being one day aliases of each other (Motch et al. 
\cite{Motch}). One of these periods is assumed to be the orbital period of 
the system.

\section{{\em ROSAT} HRI observations and results} 

\subsection{YY~Draconis}

Here we report details of the longest X-ray observation yet of YY~Dra. 
It was made with the {\em ROSAT} \ High Resolution Imager (Zombeck et al. 
\cite{Zom}) and comprises 56.6~ksec on source between 1996 May 22 23:54 UT and 
1996 May 25 06:27 UT. A lightcurve at 10~s time resolution was constructed by
using the Starlink {\em Asterix} \ software (Allen \& Vallance \cite{SUN98.6})
to optimally extract data from a region 0.68 arc min in radius, centred on the 
source. Background subtraction was carried out using the data from a 
nearby region of sky, and the resulting lightcurve is shown in Fig.~1 at a 
lower time resolution. The lightcurve is broken-up by the 90~minute 
satellite orbit and has a mean count rate of 0.40~c~s$^{-1}$. 

\begin{figure}
\setlength{\unitlength}{1cm}
\begin{picture}(7,11)
\put(0,-1){\includegraphics{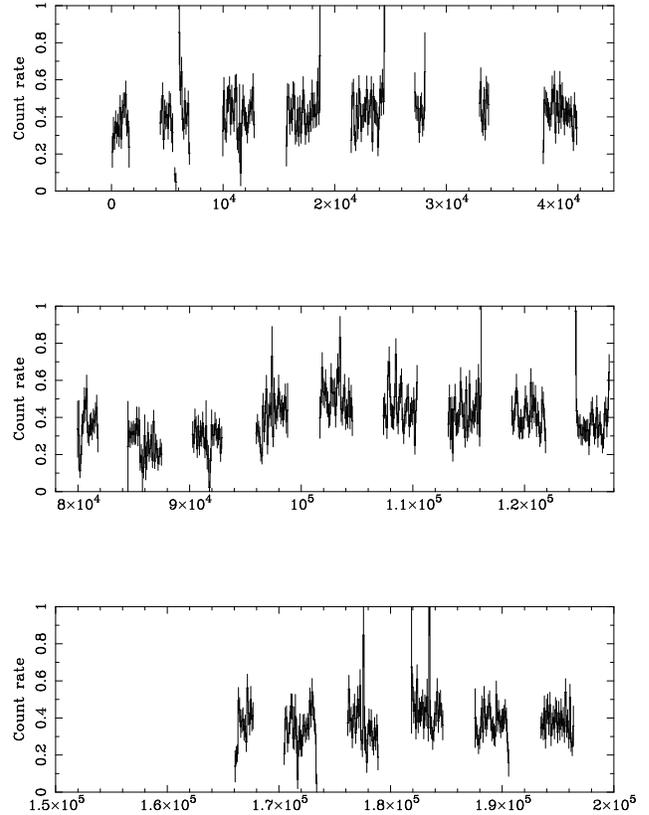}}
\end{picture}
\caption{The {\em ROSAT} HRI lightcurve of YY~Dra at a time resolution
of 100~s. There are no data in the time intervals between those illustrated
in each panel.}
\end{figure}            

To analyse the lightcurve we used the 1-dimensional {\sc clean} algorithm
in the implementation of H.J.~Lehto. This is particularly suited to 
time series which are irregularly sampled and in which multiple periodicities 
may be present. Fig.~2 shows the resulting raw power spectrum, window 
function, and {\sc clean}ed power spectrum. The only significant signal 
detected is at a frequency corresponding to a period of ($264.7 \pm 0.1$)~s, 
and may be identified with the first harmonic of the white dwarf spin 
frequency described earlier. Other spikes in the power spectrum, at a level 
of $10^{-4}$~c$^{2}$~s$^{-2}$ or lower, do not correspond to any known system
period, or combination of periods, and are presumed to be due to noise.
The power at the first harmonic of the spin frequency is 
$3.9 \times 10^{-4}$~c$^{2}$~s$^{-2}$ which corresponds to an amplitude 
of 0.040~c~s$^{-1}$. (Nb. The amplitude is equal to twice the square root
of the {\sc clean}ed power.) The modulation therefore has a fractional 
amplitude of 10\%. 

\begin{figure}
\setlength{\unitlength}{1cm}
\begin{picture}(7,11)
\put(-0.5,-2.5){\includegraphics{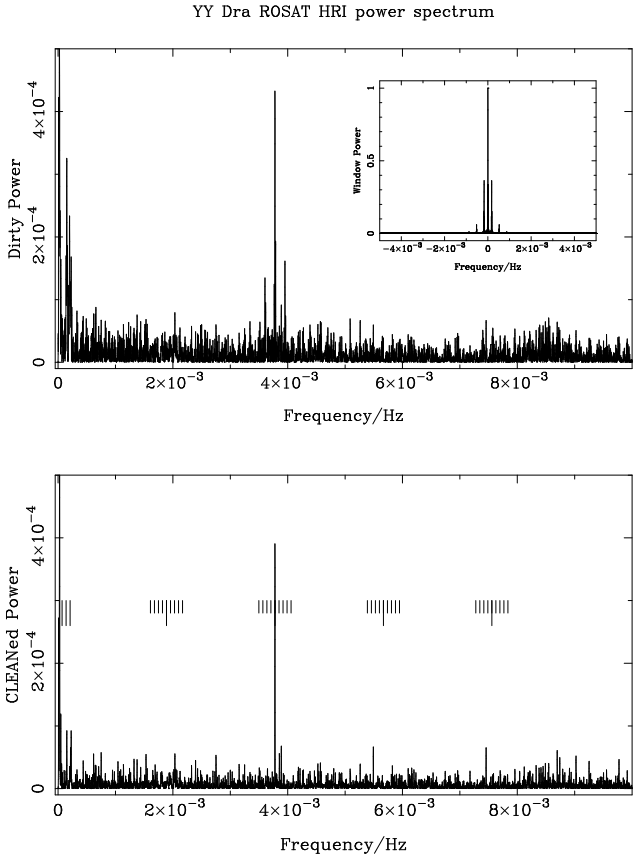}}
\end{picture}
\caption{The power spectrum of the {\em ROSAT} HRI lightcurve of YY~Dra.
The upper panel shows the raw power spectrum with the window function
inset. The lower panel shows the {\sc clean}ed power spectrum. Tick marks
show the expected locations of the orbital and spin frequencies, along with
some of their sidebands and harmonics.}
\end{figure}            

We note that no significant power is detected at either the orbital 
period or the beat period of the system, or at any harmonics and sidebands 
of either of their corresponding frequencies, in these data. In particular,
the two peaks with power of $\sim 10^{-4}$~c$^{2}$~s$^{-2}$ at frequencies 
of about $2 \times 10^{-4}$~Hz (Fig. 2) are {\em not} coincident with any 
harmonics of the orbital frequency; nor is the broad dip in the lightcurve 
around $9 \times 10^4$~s (Fig.~1) believed to be related to any orbital 
phenomenon.

The limit on the power at the actual spin period of YY~Dra (529.3~s) is 
$<10^{-5}$~c$^{2}$~s$^{-2}$, corresponding to a limiting amplitude in the 
light curve of $<0.006$~c~s$^{-1}$, and a limiting fractional amplitude
of less than 1.5\%. However, given the abundance of `noise' peaks with 
power of the order of $5 \times 10^{-5}$~c$^{2}$~s$^{-2}$, a more realistic
estimate of the limiting fractional amplitude is around 3.5\%. The data 
folded at the white dwarf spin period are shown in Fig. 3. From this, it 
is clear that the pulse profile can be described as `double-peaked' with two 
similar peaks per cycle separated by about 0.5 in phase. Since the power
at this period is clearly very small, it is not surprising that the difference 
between the two peaks is negligible.

\begin{figure}
\setlength{\unitlength}{1cm}
\begin{picture}(5,6)
\put(0,6){\includegraphics{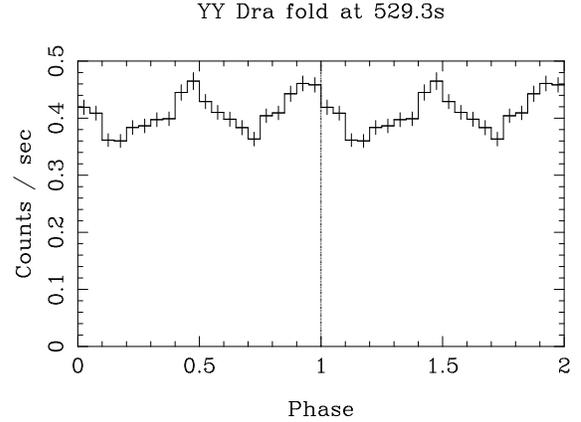}}
\end{picture}
\caption{The {\em ROSAT} HRI lightcurve of YY~Dra folded at a period 
of 529.3~s. Phase zero is arbitrary, and the profile is shown repeated 
over two cycles.}
\end{figure}

\subsection{V709~Cassiopeiae}

As with YY~Dra, we report details of the longest X-ray observation yet of 
V709~Cas, made with the {\em ROSAT} HRI. The observation comprises 43.3~ksec 
on source between 1998 Feb 15 14:31 UT and 1998 Feb 17 08:45 UT. Again using 
the Starlink {\em Asterix} \ software, data were optimally extracted from a 
region 0.65 arc min in radius, centred on the source, in order to construct a 
time series at 10~s resolution. Background subtraction was carried out using 
the data from an adjacent region of blank sky and the resulting light-curve, 
shown in Fig.~4 at a lower time resolution, has a mean count rate of 
0.26~c~s$^{-1}$. 

\begin{figure}
\setlength{\unitlength}{1cm}
\begin{picture}(7,11)
\put(0,-1){\includegraphics{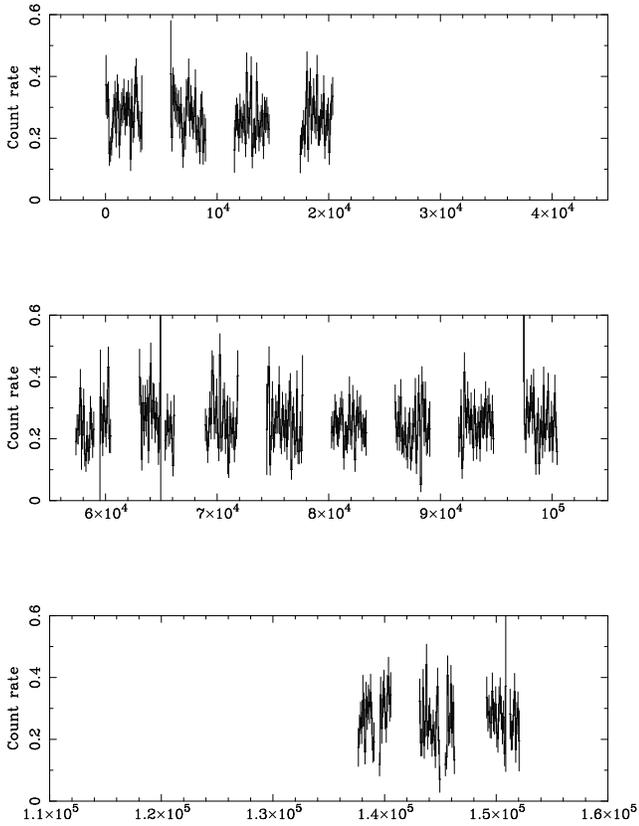}}
\end{picture}
\caption{The {\em ROSAT} HRI lightcurve of V709~Cas at a time resolution
of 100~s. There are no data in the time intervals between those illustrated
in each panel.}
\end{figure}            

As above, we used the 1-dimensional {\sc clean} algorithm to analyse the 
light-curve and Fig.~5 shows the results. We detect signals at frequencies 
corresponding to the previously identified spin period and its first and 
second harmonics. Using the accurately determined frequency of the second 
harmonic we calculate a more precise value for the spin period than previously
measured, namely $(312.78 \pm 0.03)$~s. The power at the fundamental frequency 
is $8.8 \times 10^{-4}$~c$^{2}$~s$^{-2}$ which corresponds to an amplitude 
of 0.059~c~s$^{-1}$. The modulation therefore has a fractional amplitude 
of 23\% and the data folded at a period of 312.78~s are shown in Fig.~6. 

As with YY~Dra, we note that no significant power is detected at 
either the orbital period (whether 5.4~hr or 4.45~hr) or the beat period of 
the system, or at any harmonics and sidebands of the frequencies 
corresponding to either period. The limit to the power at any other period
is of the order of $10^{-5}$~c$^{2}$~s$^{-2}$, corresponding to a limiting
fractional amplitude of about 2.5\%.

\begin{figure}
\setlength{\unitlength}{1cm}
\begin{picture}(7,11)
\put(-0.5,-2.5){\includegraphics{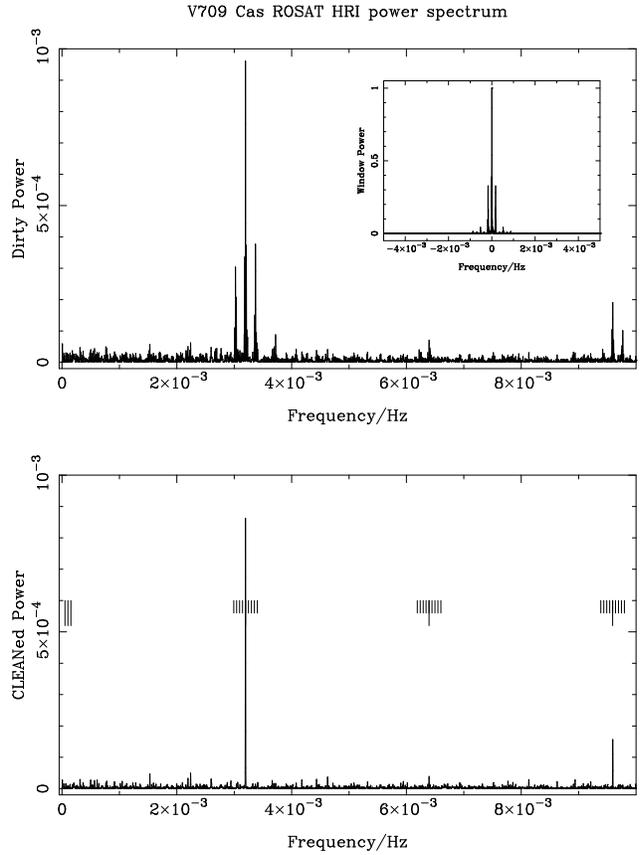}}
\end{picture}
\caption{The power spectrum of the {\em ROSAT} HRI lightcurve of V709~Cas.
The upper panel shows the raw power spectrum with the window function
inset. The lower panel shows the {\sc clean}ed power spectrum. Tick marks
show the expected locations of the spin and orbital frequencies (assuming
an orbital period of 5.4~hr), along with some of their sidebands and 
harmonics.}
\end{figure}  

\begin{figure}
\setlength{\unitlength}{1cm}
\begin{picture}(5,6)
\put(0,6){\includegraphics{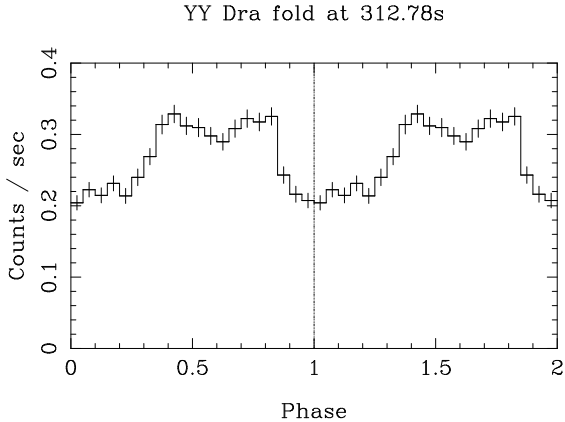}}
\end{picture}
\caption{The {\em ROSAT} HRI lightcurve of V709~Cas folded at a period
of 312.78~s. Phase zero is arbitrary and the profile is shown repeated over 
two cycles.}
\end{figure}

\section{Discussion}

\subsection{Similarities and differences}

The results from these X-ray observations of YY~Dra and V709~Cas are
clearly very similar. Both systems show a single, relatively short pulse 
period, at 264.7~s and at 312.78~s respectively, and neither system shows
any evidence for orbital or beat period modulations. The absence of an 
orbital modulation in YY~Dra is not surprising, given the relatively low 
inclination of the system, and suggests that our view of the white dwarf is 
not obscured by material located elsewhere in the binary system. The 
similar absence of orbital modulation in V709~Cas suggests that the 
inclination angle of this system too is quite low. The absence 
of a signal in the X-ray power spectrum at the beat frequency (or any related 
sideband frequency) of either system implies that both YY~Dra and V709~Cas 
must accrete predominantly via a disc rather than directly from a stream 
(Mason, Rosen \& Hellier \cite{MRH88}; Hellier \cite{Hell91}; Wynn \& King 
\cite{WK}; Norton \cite{Nor1}). 
Consequently the accreting material has lost all knowledge of the orbital 
phase by the time it impacts the white dwarf.

The similarities in pulsation behaviour between the two systems are rather 
more subtle. The spin period of the white dwarf in YY~Dra is believed to 
be {\em twice} the period observed in its power spectrum, and so the 
resulting spin pulse profile is double-peaked with two similar maxima 
(Fig. 3).  Conversely there is no evidence to suggest that the spin period 
of the white dwarf in V709~Cas is anything other than the dominant period 
seen in its power spectrum. Nonetheless we would argue that the pulse profile 
of V709~Cas (Fig. 6) is also `double-peaked'. The only difference between 
it and the pulse profile of YY~Dra is that the two maxima in the case of 
V709~Cas are separated by only about one-third of the pulse cycle, rather 
than half the pulse cycle in the case of YY~Dra. The secondary minimum in 
the V709~Cas pulse profile is therefore `filled-in' somewhat by the proximity 
of the two peaks. So, whilst the power spectrum of YY~Dra is dominated 
by the first harmonic of the spin frequency (ie. $2/P_{\rm spin}$), that 
of V709~Cas is dominated by the fundamental and the second harmonic 
(ie. $1/P_{\rm spin}$ and $3/P_{\rm spin}$). 

\subsection{Double-peaked pulse profiles as an indicator of a weak 
magnetic field}

When compared with many other intermediate polars, the `unusual' feature of 
both YY~Dra and V709~Cas is that they display double-peaked pulse profiles. 
Amongst the rest, the only other intermediate polars that have shown similar 
behaviour are AE~Aqr, DQ~Her, V405~Aur, GK~Per and XY~Ari (the latter two 
only have double peaked pulse profiles in quiescence and change to 
single-peaked pulse profiles during outburst). What these systems have in 
common is that they all have relatively short white dwarf spin periods. As
noted above, the periods of YY~Dra and V709~Cas are about 529~s and 313~s, 
whilst those of the other five systems listed are about 33~s, 142~s, 545~s, 
351~s and 206~s respectively. All other confirmed intermediate polars 
have white dwarf spin periods in excess of 700~s.

It is believed that most intermediate polars exist in a state of equilibrium
rotation (eg. Warner \cite{War96}), that is to say the accretion disc is 
disrupted at the radius where the Keplerian period of the disc is equal to 
the rotation period of the white dwarf. Now, if this is the case, then a 
short spin period implies that the white dwarf has a weak magnetic field. 
In fact, the magnetic moment of the white dwarf is proportional 
$P_{\rm spin}^{7/6}$ and it has been calculated that YY~Dra, V405~Aur and 
XY~Ari each have magnetic moments of about $10^{32}$~G~cm$^3$, for example 
(Warner \cite{War96}). So, as pointed out by Hellier (\cite{Hell96}) and 
Allan et al. (\cite{allan96}), the implication is that a short white dwarf 
spin period, and hence a weak magnetic field, is what determines the presence 
of a double-peaked pulse profile.

The common interpretation of a double-peaked spin pulse profile is to say 
that the system is undergoing two-pole accretion, and this is indeed the 
interpretation previously placed on YY~Dra (eg. Patterson \& Szkody \cite{PS}).
However, this is a simplistic interpretation since a double-peaked pulse
profile {\em is not} a unique indicator of two-pole accretion. If it were
then it would imply that single-peaked pulse profiles are the result of 
one-pole accretion. Whilst this is probably true in the phase-locked polar 
systems which undergo stream-fed accretion, it is unlikely to occur in a 
disc-fed intermediate polar system. In this case material from the inner edge 
of the disc is as likely to be channelled to one pole as the other, and 
two-pole accretion is the normal way in which such a system will accrete. 
As pointed out by Hellier (\cite{Hell96}) and Allan et al. (\cite{allan96}), 
two-pole disc-fed accretion {\em does not} generally produce a double-peaked 
pulse profile, although it can in some circumstances.  Indeed, two-pole 
disc-fed accretion is believed to be the `normal' mode of behaviour in 
intermediate polars, yet both single-peaked and double-peaked pulse profiles 
are seen. It is important that the paradigm which states `single-peaked 
profile equals one-pole accretion; double-peaked profile equals two-pole 
accretion' is put to rest for intermediate polars. We describe below how 
both types of pulse profile can be produced by two-pole disc-fed accretion, 
depending on the strength of the magnetic field.

\subsection{Two-pole disc-fed accretion producing a single-peaked pulse 
profile}

Many intermediate polars, such as the canonical system AO~Psc (Hellier, 
Cropper \& Mason \cite{HCM}), show a single-peaked pulse profile resulting 
from two-pole disc-fed accretion. With a relatively strong magnetic field, 
the accreting material attaches to the field lines whilst still quite distant 
from the white dwarf, as shown in Fig. 7. This results in relatively small 
emission regions, whose `vertical optical depth' (up the accretion curtain,
along the magnetic field lines) is greater than their `horizontal 
optical depth' (across the accretion curtain, parallel to the white dwarf 
surface). So in this case minimum attenuation of the X-ray flux (minimum 
photoelectric absorption and electron scattering) occurs when the emission 
region is seen from the side. Hence, when the `upper pole' (ie. the one in 
the hemisphere of the white dwarf above the orbital plane as we observe it) 
points away from the observer, then pulse maximum is seen (Fig. 7 viewpoint 
B), and when it points towards the observer, pulse minimum is seen (Fig. 7 
viewpoint A). The contribution to the modulation from the lower pole is 
{\em in phase} with that from the upper pole, since when the upper pole is 
pointing towards the observer, the lower pole will generally be occulted. 
Conversely, when the upper pole is pointing away from the observer, the lower 
pole is viewed essentially from the side too, and so its flux adds to the 
pulse giving a maximum. 

Single-peaked, roughly sinusoidal, pulse-profiles are seen in many 
intermediate polars, and so two-pole disc-fed accretion can be considered 
as the `normal' mode of behaviour in these systems.

\begin{figure}
\setlength{\unitlength}{1cm}
\begin{picture}(6,7)
\put(-4,-8){\includegraphics{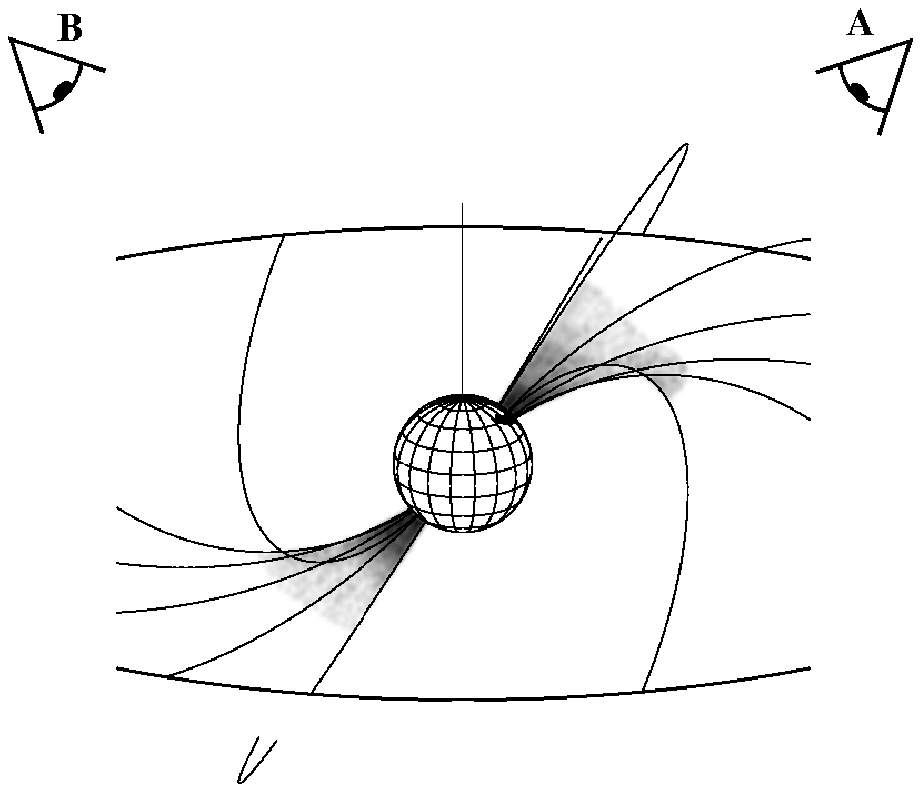}}
\end{picture}
\caption{Schematic diagram showing an intermediate polar with a relatively
high magnetic field. Attenuation of the X-rays is greatest along the magnetic 
field lines (up the accretion curtains), and least parallel to the white
dwarf surface (across the accretion curtains). When viewed from B, X-ray 
pulse maximum is seen, and when viewed from A, X-ray pulse minimum is seen.}
\end{figure}

\subsection{Two-pole disc-fed accretion producing a double-peaked pulse 
profile}

In contrast to the above, Hellier (\cite{Hell96}) and Allan et al. 
(\cite{allan96}) suggest that a double-peaked pulse profile can result if 
the vertical optical depth (up the accretion curtain) is {\em lower} than 
the horizontal optical depth (across it). As they point out, if the white 
dwarfs in systems showing a double-peaked pulse profile have a relatively 
weak magnetic field, then material threads onto the field lines much closer 
to the white dwarf, and the accretion area is relatively large, as shown in 
Fig.~8. With a large enough footprint to the accretion curtain, this optical 
depth reversal will be inevitable. This is reminiscent of the large accretion 
area model suggested by Norton \& Watson (\cite{nw}).

Now when the upper pole points towards the observer, maximum flux is seen 
from it, so giving the first peak in the pulse profile (Fig. 8 viewpoint A).
The contribution to the modulation from the lower pole is {\em in anti-phase} 
with that from the upper pole since, when the upper pole is pointing towards 
the observer, the lower pole will generally be occulted, and when the upper 
pole is pointing away from the observer, the lower pole is at its most 
visible, so giving a second peak in the pulse profile (Fig. 8 viewpoint B). 
A relatively large accretion area may also account for the fact that emission 
from the lower pole is seen even at the relatively low inclination angle of 
$45^{\circ}$ in YY~Dra.

\begin{figure}
\setlength{\unitlength}{1cm}
\begin{picture}(6,7)
\put(-4,-8){\includegraphics{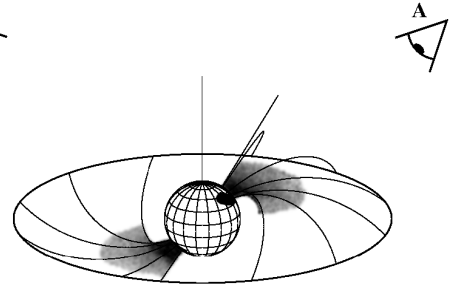}}
\end{picture}
\caption{Schematic diagram showing an intermediate polar with a relatively
low magnetic field. Attenuation of the X-rays is greatest parallel to the
white dwarf surface (across the accretion curtains), and least along the 
magnetic field lines (up the accretion curtains). X-ray pulse maxima are seen
when the system is view from both A and B.}
\end{figure}

An alternative way of producing a double-peaked pulse profile from two-pole
accretion may be to have {\em tall} accretion regions, and this could also 
conceivably be the result of a weak magnetic field. The shock height 
is proportional to the size of the accreting area (eg. Frank, King \& Raine 
\cite{FKR}), so the likelihood is that the accretion regions in intermediate 
polars with a weak magnetic field are both wide and tall. Even if the accretion
region is not wide enough for the vertical optical depth to be lower than
the horizontal optical depth, it may be tall enough for each accretion 
region to not entirely disappear over the rim of the white dwarf. 
With tall accretion regions whose vertical optical depths are greater than 
their horizontal optical depths (as in a conventional 
accretion curtain) therefore, when the upper pole points away from the 
observer, the lower pole is viewed essentially from the side, and so its flux 
adds in phase to the first flux maximum of the cycle. But, when the upper 
pole points towards the observer (giving minimum flux), the lower pole 
{\em may still be visible} and so give rise to a second flux maximum in 
the cycle. 

Having said this, the height of the accretion region also
depends inversely on the mass accretion rate. In most intermediate polars,
the mass accretion rate is probably not low enough for the accretion region
to have an appreciable height and so make this scenario feasible. The one
exception is probably \object{EX Hya} which has an accretion rate about ten 
times lower than most other intermediate polars, and consequently an accretion
shock height of about one white dwarf radius. In that case though, the upper 
pole is continuously visible and partial occultation of the lower pole is the
main cause of the single-peaked spin pulse profile (Allan, Hellier \& 
Beardmore \cite{allan98}). 

In summary, either wide or tall accretion regions could give rise to a 
double-peaked pulse profile, and both could be the result of a relatively
weak white dwarf magnetic field. Such double-peaked pulse profiles
might therefore be expected to arise in intermediate polars whose white
dwarfs have relatively short spin periods.

\subsection{A subset of intermediate polars with weak magnetic fields}

Allan et al. (\cite{allan96}) discussed the cases of V405~Aur, AE~Aqr and 
YY~Dra in support of the theory outlined above, whilst Hellier (\cite{Hell96})
also mentioned DQ~Her and XY~Ari in addition to those three. As further 
support for this theory we can now add V709~Cas to the list of sources and 
we also note that GK~Per, with a spin period of 351~s, has displayed a 
double-peaked pulse profile on some of the occasions it has been observed. 

The data presented earlier demonstrate that V709~Cas follows the pattern as 
an intermediate polar with a short spin period displaying a double-peaked
pulse profile. The fact that the two maxima in its pulse profile appear more 
closely spaced than those of YY~Dra is most probably due to an asymmetry in 
the locations of the upper and lower magnetic poles. This may be caused by a 
dipole magnetic field which is offset from the centre of the white dwarf, for 
instance. If the two poles are not separated by 180$^{\circ}$, then the 
times at which maximum flux is seen from the two poles will not be separated 
by 0.5 in phase either. 

GK~Per only exhibits its double-peaked pulse profile in quiescence (Norton, 
Watson \& King \cite{nwk}; Ishida et al. \cite{Ish}), and shows a 
single-peaked pulse profile when in outburst (Watson, King \& Osborne 
\cite{WKO}). A similar situation exists in XY~Ari, which also shows a 
double-peaked pulse profile when the system is in quiescence ({\em Ginga} 
observation: Kamata \& Koyama \cite{KK93}; {\em RXTE} observation: Hellier 
\cite{Hell97a}), but a single-peaked pulse profile when in outburst (Hellier, 
Mukai \& Beardmore \cite{Hell97b}). We suggest that the cause of the 
changing pulse profile may be the same in both cases, and can be understood 
in the light of the model outlined earlier, following the explanation given
by Hellier et al. (\cite{Hell97b}). The radius at which the accretion 
disc is truncated is proportional to (amongst other things) $\dot{M}^{-2/7}$ 
(eg. Frank, King \& Raine \cite{FKR}). Since the disc already extends fairly 
close to the white dwarf in both these systems, due to the relatively weak 
magnetic field, during outburst the increased mass accretion rate will cause 
the accretion disc to extend even closer to the white dwarf before it is 
truncated. Hellier et al. (\cite{Hell97b}) suggest that in XY~Ari the 
accretion disc then hides the lower pole from view, and a single-peaked pulse 
profile remains, produced by modulation of the X-ray flux from the upper pole 
only. Now, XY~Ari is an eclipsing system, with $i=82^{\circ}$ (Hellier 
\cite{Hell97a}), whilst the inclination of GK~Per is believed to be within 
the range $46^{\circ}<i<73^{\circ}$ (Reinsch \cite{Reinsch94}). In XY~Ari the 
accretion disc extends to within four white dwarf radii of the white dwarf 
during outburst (Hellier et al. \cite{Hell97b}). So, if the lower pole in 
GK~Per is also hidden during outburst, the implication is that the accretion 
disc must extend even closer to the white dwarf in that case.  

If the model outlined in Sect. 4.4 is correct, then we would expect 
{\em all} short period intermediate polars to display double-peaked pulse 
profiles. The seven systems described above comprise the only {\em confirmed} 
intermediate polars with a spin period below about 700~s. However, there are 
also three systems that have been proposed as intermediate polars which would 
fall within this subset if their classifications are confirmed.

The first of these systems is \object{V533 Her}, and although it has never 
exhibited X-ray pulsations, it has been suggested that it is an intermediate 
polar on the basis of optical photometry which shows a stable 63~s period 
that appears and disappears with a timescale of years (Patterson \cite{Patt}).
As a short period, weak magnetic field system, we might expect its X-ray spin 
pulse profile to be double-peaked, if such a pulsation is ever detected. In 
this case the previously identified pulse period may in fact represent half 
the true spin period of the white dwarf. A similar re-assessment of the spin 
period of DQ~Her has recently been made (Zhang et al. \cite{Zhang}) resulting 
in the identification of its spin period as 142~s, twice the value previously 
assumed.

The other two proposed intermediate polars are both recently discovered
systems by {\em ROSAT}. \object{RX J0757.0+6306} has an optical photometric 
modulation with a period of 511~s which may represent the spin period of the 
white dwarf (Tovmassian et al. \cite{Tov98}). However, no X-ray pulsation 
at this period has been detected and the classification as an intermediate 
polar has yet to be confirmed. \object{RX J1914.4+2456} displays an X-ray 
`pulsation' with a period of 569~s, but Cropper et al. (\cite{Cropp98}) 
suggest that this may be a double degenerate polar, rather than an 
intermediate polar, and so the period represents the orbital period of the 
system rather than the spin period of a white dwarf. If either of these 
systems do turn out to be intermediate polars, we predict that their pulse 
profiles may turn out to be double-peaked also.

\subsection{Beat frequency signals}

None of the seven systems that show double-peaked pulse profiles has shown 
evidence for beat frequency signals in their X-ray emission. Such a 
signal is generally taken as a signature of stream-fed or disc-overflow 
accretion, and may therefore be confined to the intermediate polars with 
higher magnetic field strengths. This is to be expected, as the accretion 
flow becomes attached to the magnetic field lines at larger distances from 
the white dwarf when the magnetic field is stronger. The further from the 
white dwarf, the more chance there is that some of the accretion flow is 
still constrained to travel as an accretion stream, so the greater the 
likelihood of a signature of stream-fed accretion in the X-ray power 
spectrum. 

All the intermediate polars that have exhibited X-ray beat period
signals have relatively long white dwarf spin periods. FO~Aqr with a spin
period of 1254~s showed a strong beat period during a {\em Ginga} \ observation
in 1988 and in an {\em ASCA} \ observation in 1993, although it was absent 
during a second intervening {\em Ginga} \ observation in 1990 (Norton et al. 
\cite{nort92a}; Beardmore et al. \cite{beard98}). TX~Col with a spin period 
of 1911~s showed a strong beat period in an {\em EXOSAT} \ observation in 
1985 and also in a {\em ROSAT} \ HRI observation from 1995, however it was 
absent from an {\em ASCA} \ observation in 1994 (Buckley \& Tuohy 
\cite{Buck89}; Norton et al. \cite{nort97}). The case of BG~CMi is more 
controversial. Here an 847~s period detected during a 1988 {\em Ginga} \ 
observation was interpreted by Norton et al. (\cite{nort92b}) as the 
{\em true} spin period, implying that the previously detected 
913~s X-ray pulsation was at the beat period. Even more radically, Patterson 
\& Thomas (\cite{PT}) have suggested that the true spin period is 1693~s, 
with the 913~s X-ray pulse then corresponding to a frequency of 
$(2/P_{\rm spin} - 1/P_{\rm orbit})$. In either interpretation the strong 
X-ray pulse is an indicator of a stream-fed accretion component, but the case 
for this is not yet proved one way or the other. The two systems AO~Psc and 
V1223~Sgr, with spin periods of 805~s and 745~s respectively, both showed 
tentative evidence for beat periods in {\em EXOSAT} \ data from 1983--1985 
(Hellier \cite{Hell92}), although later observations with {\em Ginga}, 
{\em ROSAT} \ and {\em ASCA} \ failed to detect any (Taylor et al. 
\cite{Tay97}; Hellier et al. \cite{hemu96}). Finally, 
the recently discovered intermediate polar RX~J1712.6--2414 displays an 
X-ray pulsation only at the beat period of 1003~s, with no signal at the 
927~s spin period of the white dwarf (Buckley et al. \cite{Buck97}). 
Furthermore, RX~J1712.6--2414 and BG~CMi are two of the three intermediate 
polars from which polarized emission has been detected (Buckley et al. 
\cite{Buck95}; Penning, Schmidt \& Liebert \cite{Penn86}; West, Berriman 
\& Schmidt \cite{WBS}), which is another 
signature of a strong magnetic field. The correlation between X-ray beat 
periods, strong magnetic fields and long white dwarf spin periods is clearly 
apparent. 

We therefore suggest that other intermediate polars with 
long white dwarf spin periods and strong magnetic fields might be expected
to exhibit X-ray beat periods at some time in their lives. A prime 
candidate to search for such effects may be PQ~Gem with a spin period 
of 833~s (Mason et al. \cite{Mason92}), since this is the third system for 
which polarized emission has been detected (Piirola, Hakala \& Coyne 
\cite{Pii93}). Observations with {\em Ginga} \ and the {\em ROSAT} \ PSPC 
failed to detect an X-ray beat period signal, but the upper limits to the 
amplitudes of such a modulation were 5\% and 20\% respectively (Duck et al. 
\cite{Duck94}), so it cannot 
be ruled out. Moreover, as the observations of FO~Aqr and TX~Col demonstrate,
beat period signals can appear or disappear on a timescale of a few years,
so may yet be found in PQ~Gem.

\section{Summary}

We have presented data obtained with the {\em ROSAT} HRI which constitute 
the most sensitive X-ray observations yet of the intermediate polars YY~Dra 
and V709~Cas. In common with some previous observations of YY~Dra, the power 
spectrum of its X-ray lightcurve shows a signal only at a frequency 
corresponding to 264.7~s, which is {\em half} the spin period of the white 
dwarf. The X-ray spin pulse, with a period of 529.3~s, therefore exhibits a 
double-peaked profile in which the two peaks are similar and separated
by about 0.5 in phase. The power spectrum of the X-ray lightcurve of V709~Cas 
shows a signal at a frequency corresponding to 312.8~s, the previously 
determined spin period, and also at the first two harmonics of this 
frequency. The harmonics enabled us to measure the spin period more accurately 
as $(312.78 \pm 0.03)$~s. When folded at this period, the pulse profile 
of V709~Cas displays a structure that is also double-peaked, 
although the maxima are separated by only about 0.3 in phase, and the 
secondary minimum is somewhat filled-in. There is no evidence for X-ray 
modulation at either the orbital period or the beat period of either system, 
so each must be disc-fed and seen at a relatively low inclination angle.

To allay a common assumption, we have emphasised that a double-peaked pulse 
profile is {\em not} a unique indicator of two pole accretion, and that
either single-peaked {\em or} double-peaked pulse profiles can arise when 
accretion occurs onto both poles of the white dwarf. Following Hellier 
(\cite{Hell96}) and Allan et al. (\cite{allan96}), we have explained that the 
two possibilities are 
related to the spin period of the white dwarf. A short white dwarf spin 
period is an indicator of a relatively weak magnetic field, which in turn 
gives rise to relatively large footprints of the accretion curtains on the 
white dwarf surface. Conversely, a long white dwarf spin period indicates a 
somewhat stronger magnetic field, which means that the footprints of the 
accretion curtains are smaller. In the first case the relative optical 
depths across and along the accretion curtain conspire to produce a 
double-peaked pulse profile, whilst in the second case a single-peaked 
pulse profile is the result. We trust that this will lay to rest the widely
accepted paradigm that single-peaked X-ray pulse profiles are the result of
single pole accretion whilst double-peaked X-ray pulse profiles are the only
result of two pole accretion.

There are now seven intermediate polars with short spin periods which display 
double-peaked pulse profiles and must therefore have weak magnetic fields: 
AE~Aqr, DQ~Her, XY~Ari, V709~Cas, GK~Per, YY~Dra and V405~Aur.
None of them exhibit X-ray beat periods, so stream-fed or disc overflow 
accretion does not occur in these systems. Conversely, the six systems
which have shown X-ray beat periods at some time all have long spin  
periods and must have stronger magnetic fields: FO~Aqr, TX~Col, BG~CMi,
AO~Psc, V1223~Sgr and RX~J1712.6--2414. We conclude that double peaked 
X-ray pulse profiles are an indicator of a white dwarf with a relatively weak 
magnetic field, whilst X-ray beat periods are an indicator of a white dwarf 
with a relatively strong magnetic field.

\begin{acknowledgements}
The data analysis reported here was carried out using facilities provided 
by PPARC, Starlink and the Open University Research Committee. 
\end{acknowledgements}

\end{document}